\documentclass[aps,prb,twocolumn,superscriptaddress]{revtex4-2}
\usepackage[english]{babel}
\usepackage{bm}
\usepackage{amsmath}
\usepackage{graphicx}
\usepackage{dcolumn}% Align table columns on decimal point
\usepackage[colorlinks=true, allcolors=blue]{hyperref}
\usepackage{diagbox}
\usepackage{multirow}
\usepackage{threeparttable}

\begin{document}

\title{Scaleability of dielectric susceptibility  $\epsilon_{zz}$ with the number of layers and additivity of ferroelectric polarization in van der Waals semiconductors}

\author{F. Ferreira}
\affiliation{University of Manchester, School of Physics and Astronomy, Oxford Road, Manchester M13 9PL, United Kingdom}
\affiliation{National Graphene Institute, University of Manchester, Booth St. E. Manchester M13 9PL, United Kingdom}

\author{V.~V.~Enaldiev}
\affiliation{University of Manchester, School of Physics and Astronomy, Oxford Road, Manchester M13 9PL, United Kingdom}
\affiliation{National Graphene Institute, University of Manchester, Booth St. E. Manchester M13 9PL, United Kingdom}

\author{V.I. Fal'ko}
\affiliation{University of Manchester, School of Physics and Astronomy, Oxford Road, Manchester M13 9PL, United Kingdom}
\affiliation{National Graphene Institute, University of Manchester, Booth St. E. Manchester M13 9PL, United Kingdom}
\affiliation{Henry Royce Institute, University of Manchester, Booth St. E. Manchester M13 9PL, United Kingdom}

\begin{abstract}
    We study the dielectric response of few layered crystals of various transition metal dichalcogenides (TMDs) and hexagonal Boron Nitride (hBN). 
    We showed that the out-of-plane polarizability  of a multilayer crystal (which characterizes response to the external displacement field)   scales linearly with the number of layers, $\alpha_{zz}^{NL} =N  \alpha_{zz}^{1L}$, independently of the stacking
    configuration in the film. 
    We also established additivity of ferroelectric polarizations of consecutive interfaces in case when such interfaces have broken inversion symmetry.
    Then we used the obtained data of monolayer $\alpha_{zz}^{1L}$ to calculate the values of the dielectric susceptibilities for semiconductor TMDs and hBN bulk crystals. 
\end{abstract}

\maketitle

\section{Introduction}

Dielectric permittivity is an important parameter for modeling optoelectronic devices.
It characterizes how a material is polarized under an external electric field, which is relevant for modeling  field-effect transistors \cite{Younis2021,Sebastian2021}, capacitors \cite{Poonam2019}, and ferroelectrics based memristors \cite{Mikolajick2021,Weston2022}. 
In layered materials, dielectric permittivity reflects\cite{Huser2013,Latini2015,Huser2013,Ferreira2019,Ceferino2020} a strong anisotropy of crystalline and electronic properties, which is particularly strong in van der Waals (vdW) layered crystals such as graphite, black phosphorus, hexagonal boron nitride (hBN), and transition metal dichalcogenides (TMDs).
Because of the layered nature of these compounds, all of them had already been implemented as components in various field-effect transistor devices, where electrostatics is determined by the out-of-plane component of the dielectric permitivitty tensor, $\epsilon_{zz}$.
Despite its importance for device modeling, only few theoretical studies have been dedicated to the evaluation of $\epsilon_{zz}$ in layered vdW semiconductors such as InSe, GaSe, MoS$_2$, WS$_2$, MoSe$_2$, WSe$_2$, or MoTe$_2$, and these published\cite{Geick1966, Wieting1971, Neville1976, Ohba2001, Cai2007, Ghosh2013, Lu2014,  Liang2014, Kim2015, Wang2016,  Koo2017, Pike2018, Farkous2019,  Laturia2020} broadly disagree on their values and even qualitative dependence on the material thickness.

\begin{figure*}
\includegraphics[width=1.0\linewidth ]{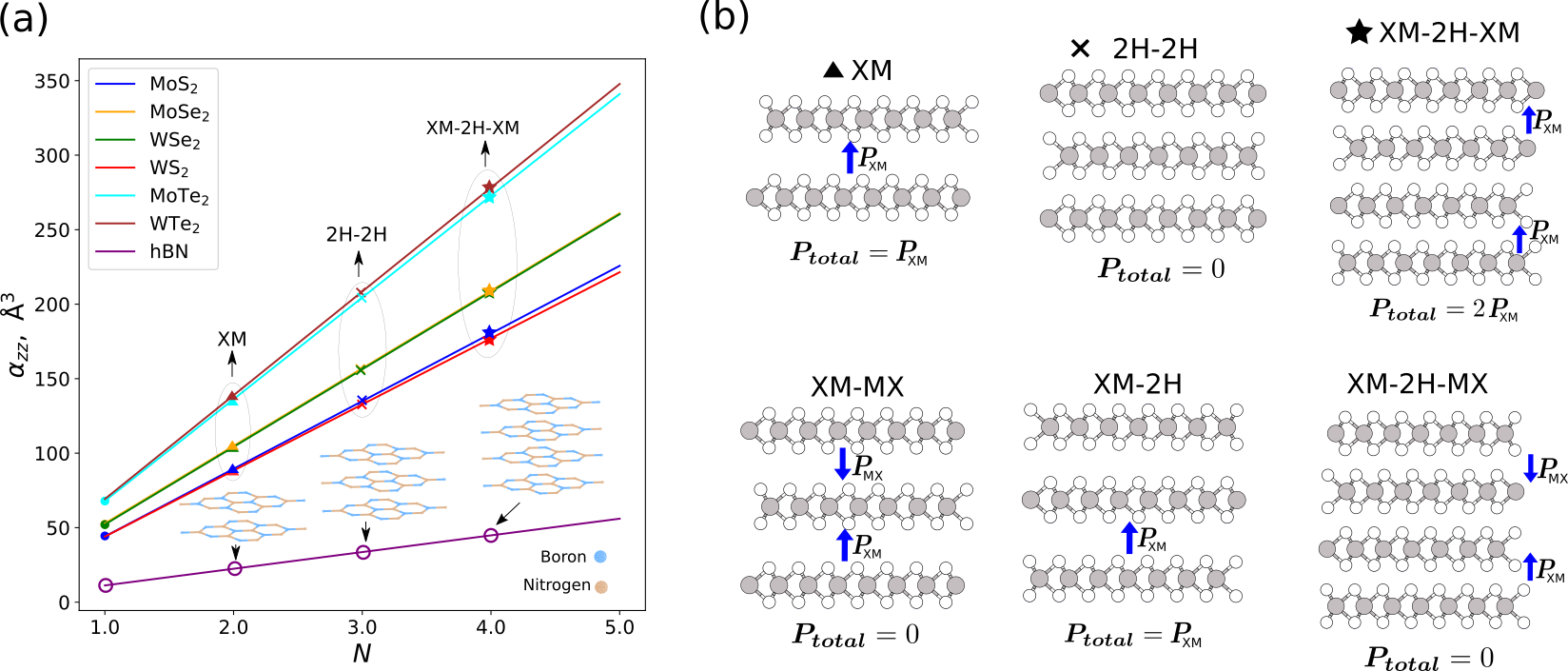}
\caption{(a) Out-of-plane polarizability ($\alpha_{zz}$) as a function of the number of layers $N$.
Linear regressions for all materials show a coefficient of determination of $R^2>0.999$.
Symbols correspond to values gathered from DFT calculations and lines correspond to fittings via a linear regression. 
The geometries that were used for multilayer TMDs in panel (a) are represented in panel (b), where it is illustrated the directions of the ferroelectric polarizations,  $\boldsymbol{P}_{\mathrm{MX}}$ and $\boldsymbol{P}_{\mathrm{XM}}$,  and how they contribute to the total ferroelectric polarization $\boldsymbol{P}_{total}$.
Configurations used for multilayer hBN crystals can be seen in panel (a).} 
\label{Fig:alpha_vs_Nlayers}  
\end{figure*}

Here, we perform a detailed ab initio density functional theory (DFT) study of $\epsilon_{zz}$ in few-layer films of MX$_2$ (M=Mo,W and X=S,Se,Te) and hBN. 
We  compute the out-of-plane polarizability, $\alpha_{zz}$, of crystals with  different numbers of layers and established linear scaleability of $\alpha_{zz}$ with the number of layers, as illustrated in Fig. \ref{Fig:alpha_vs_Nlayers}(a).
This indicates that each layer screens the external electric field independently, in agreement with other works\cite{Yu2008, Koo2017,  Tian2020}. 
In this study we take into account both  electronic and ionic polarizabilities which enables us to establish the values of static and high frequency (higher than optical frequency) values of $\epsilon_{zz}$, which appears to be particular important for hBN.
We also model various stacking arrangements of layers in the multilayers, e.g., as shown in Fig  \ref{Fig:alpha_vs_Nlayers}(b), in particular  those that allow for inversion symmetry broken interfaces. 
For such interfaces, the ferroelectric polarization is possible due to the interlayer charge transfer which also showed to be additive for consecutive interfaces.

Below, in section II,  we start by discussing the DFT method which enables us to compute the values of  $\alpha_{zz}$ for monolayers and multilayers taking into account that some of those exhibit ferroelectric interfaces.
These results presented in section II demonstrate scaleability of  $\alpha_{zz}$ in semiconducting TMDs and multilayer hBN.
In section III the computed values of  $\alpha_{zz}$ are recalculated into $\epsilon_{zz}$ of a bulk crystal which appears to be a parameter independent of the number of layers in the slab.

\section{Computation of polarizability $\alpha_{zz}$ }

In this work the method of choice is to compute the dependence of the energy of a thin slab of a crystal subjected to an out-of-plane displacement field using DFT. 
In this calculation a displacement field enters via gradient of a sawtooth potential imposed onto periodically placed  few-layer 2D materials with a large spacer along $z$-axis. 
The external  displacement field induces a dipole  moment $\alpha_{zz} D$ which screens the displacement field inside the film and determines the material dielectric constant as it will be discussed in the next section.
Here we describe the results of DFT calculations of  $\alpha_{zz}$ for monolayer, bilayer, trilayer and tetralayer crystals from compounds listed in the introduction.

%%%%%%%%%%%%%%%%%%%%%%%%%%%%%%%%%%%%%%%%%%%%%%
\begin{figure*}%[!h]
\includegraphics[width=1.0\linewidth ]{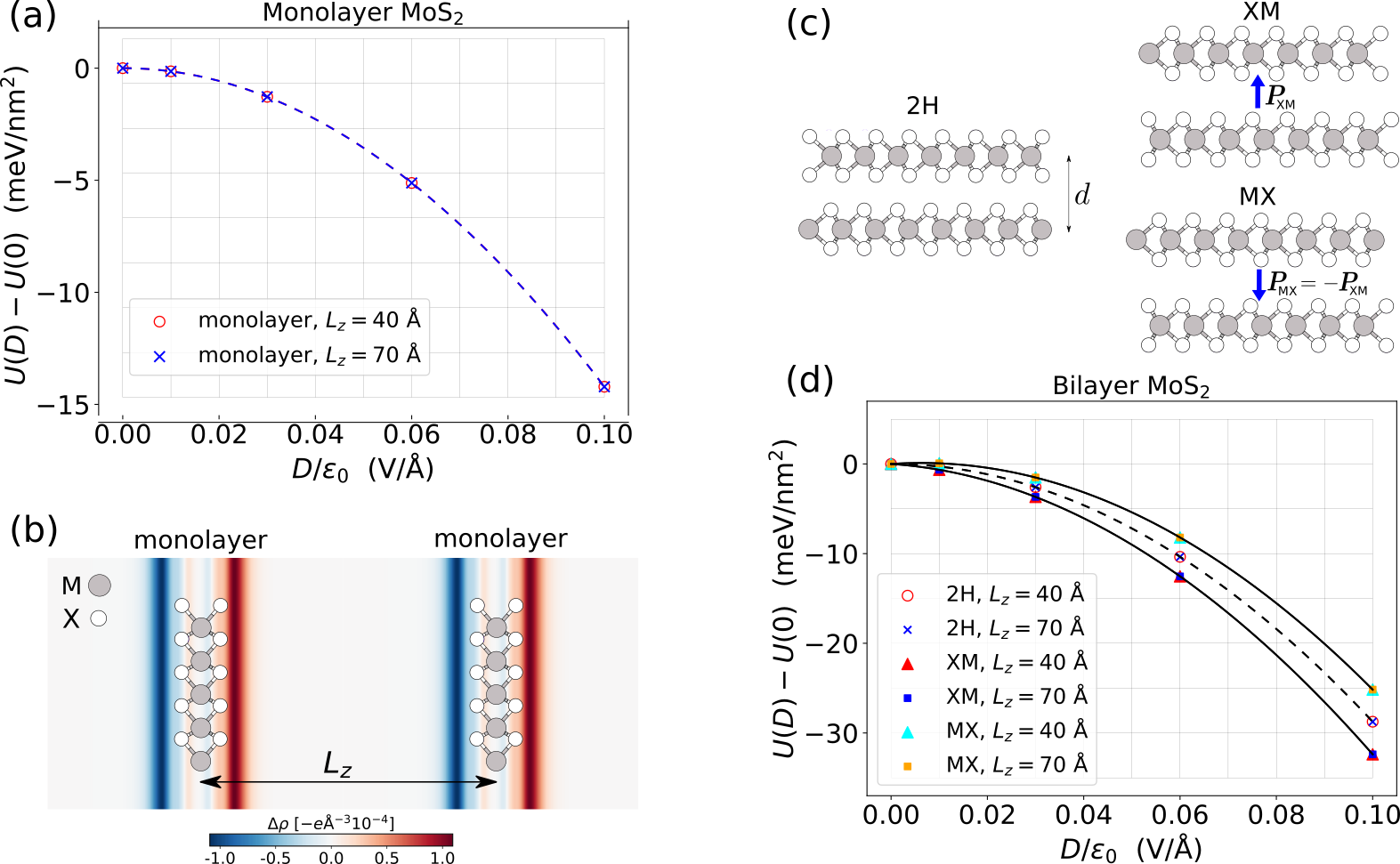}
\caption{ Total energy per unit cell area ($U$) of a monolayer of MoS$_2$ (a) and bilayer of MoS$_2$ with 2H-, XM- and MX-stacking configurations (b) as a function of  an external out-of-plane displacement field ($D$).   $U(0)$ is the total energy per unit cell area for a $D=0$. 
Symbols represent DFT data, whereas dashed and solid lines represent the fittings to DFT data, which were computed by using the expression in Eq. (\ref{eq:total_energy}) and (\ref{eq:total_energy_bilyer}) for monolayer and bilayers respectively. 
Different out-of-plane periodicities ($L_{z}$) between the crystals were used in our calculations. 
They are well converged at  $L_{z} = 40\  \mathrm{\AA}$ for the corresponding systems. 
The out-of-plane averaged charge density difference for a monolayer is shown in panel (b), where blue and red regions represent the accumulation and depletion of electrons, respectively.
The different stacking configurations used for our bilayer TMDs are shown in panel (c), where $d$ is the interlayer distance and we show the directions of the ferroelectric polarization $\boldsymbol{P}_{\mathrm{MX}}$ and $\boldsymbol{P}_{\mathrm{XM}}$ due to interlayer charge transfer.
The parameters obtained from our fittings are gathered in Table \ref{tab_fittings}.
}
\label{Fig:Energy_versus_Displacement_field_sl_MoS2}  
\end{figure*}
%%%%%%%%%%%%%%%
%%%%
%%%%
\begin{figure*}
\includegraphics[width=1.0\linewidth ]{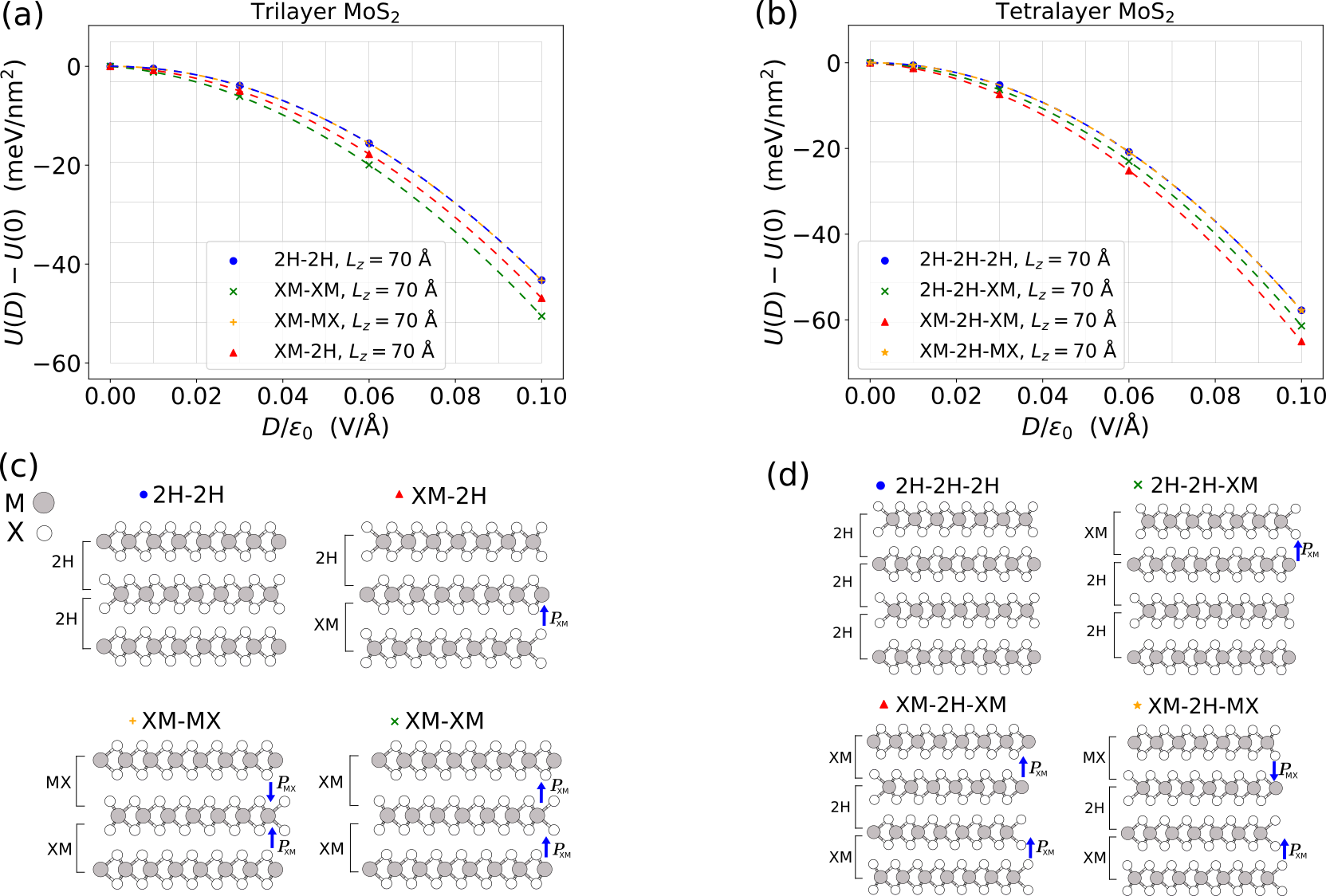}
\caption{ Total energy per unit cell area ($U$) of a trilayer (a) and tetralayer of MoS$_2$ (d) composed of different interfaces as a function of  an external out-of-plane  displacement field ($D$).
Symbols represent DFT data, whereas dashed and solid lines represent the fittings to DFT data, which were computed by using the expression in Eq. (\ref{eq:total_energy_bilyer}).
Our results are well converged for an out-of-plane periodicity of $L_{z} = 70\  \mathrm{\AA}$.
The interfaces used for trilayers and tetralayers TMDs are shown in panels (c) and (d) respectively and we show the directions of the ferroelectric polarization $\boldsymbol{P}_{\mathrm{MX}}$ and $\boldsymbol{P}_{\mathrm{XM}}$ due to interlayer charge transfer.
The parameters obtained from our fittings are gathered in Table \ref{tab_fittings}.
}
\label{Fig:Energy_versus_Displacement_field_3l_4l_MoS2}  
\end{figure*}
%%%%
%%%%

\subsection{TMD monolayers}

Here we use the approach implemented earlier in the analyses of the out-of-plane polarizability of a monolayer graphene\cite{Slizovskiy2021}.
For this we compute the total energy of a 2D crystal per unit cell area ($U$) as a function of  the out-of-plane displacement field ($D$) and fit it with a parabolic dependence\cite{Zbigniew2017},
\begin{equation}
U = U_0  - \frac{1}{2}\frac{\alpha_{zz}^{1L} D^2 }{ \epsilon_0 \mathcal{A}}.
\label{eq:total_energy}
\end{equation}
Here $U_0$ is the energy of a unit-cell for $D=0$,  $\mathcal{A}$ is the area of the unit-cell, and $\epsilon_0$ is the permittivity of vacuum.
The DFT calculations were carried out using the plane-wave code implemented in Quantum Espresso package \cite{QE1,QE2}. 
A plane wave cut-off of 70 Ry was used for all calculations  with TMDs, where the integration over the Brillouin zone was performed
using the scheme proposed by Monkhorst-Pack \cite{kgrid} with a grid of $13 \times 13\times 1$.
We used full relativistic ultrasoft pseudopotentials with spin-orbit interaction and an exchange correlation functional that is approximated by using the PBE method \cite{PBE}.
The convergence threshold for self-consistency was  set to $10^{-9}$ Ry. 
A Coulomb truncation\cite{CoulombCut} in the out-of-plane direction  was used for all calculations and the displacement field was implemented with a $z$-dependent sawtooth potential. 
Calculations that allowed relaxation of the atomic coordinates were done using the BFGS quasi-newton algorithm, where the atoms were relaxed until the total force acting on them was smaller than $10^{-5}$  Ry/Bohr.

The above described DFT modeling was implemented in three ways.
In one we used frozen lattice positions of all ions with spacings set to values from earlier literature\cite{schutte1987crystal,bronsema1986structure}.
In the second calculation we relaxed the lattice positions of the ions for $D=0$ imposing the lattice constant fixed to the experimentally known value without allowing  further relaxation at a finite $D$.
These two calculations returned the values of $\alpha_{zz}$ which can be attributed to a purely electronic response to the external perturbation which will enable us to describe the dielectric permittivity at frequencies higher than the optical phonon frequencies.
In the third calculation we implemented lattice relaxation (still fixing the lateral lattice constant) for all values of $D$, which gives us a combined electron and ionic polarizabilities and enables us to describe static ($\omega = 0$) susceptibility of the crystal.

In Fig. \ref{Fig:Energy_versus_Displacement_field_sl_MoS2}(a),  we show a typical dependence of $U(D)$ exemplified for MoS$_2$. 
To demonstrate convergence of the result against the spacing between the $z$-direction we show data in Fig. \ref{Fig:Energy_versus_Displacement_field_sl_MoS2}(a) for two calculations outputs: circles for an out-of-plane period of $40\  \mathrm{\AA}$ and crosses for $70\  \mathrm{\AA}$.
The data in Fig. \ref{Fig:Energy_versus_Displacement_field_sl_MoS2}(a) corresponds to $\alpha_{zz}^{1L} = 44.46\  \mathrm{\AA^3}$.
In Fig. \ref{Fig:Energy_versus_Displacement_field_sl_MoS2}(b) we show the distribution of $\Delta \rho_{zz}= \rho_{zz}^{D  \neq 0 } - \rho_{zz}^{D = 0}$ averaged over the unit-cell of a TMD, which indicates that polarizability is dominated by the contribution of chalcogen orbitals.
The values of $\alpha_{zz}$ computed in the above-described three ways are gathered in Table \ref{tab:polarizability_elec_ionic} for various TMDs.
The comparison of the last two columns of Table \ref{tab:polarizability_elec_ionic}  indicates that ionic contribution towards TMD polarizability is less than $0.2 \%$.
Therefore in the following analysis of few-layered crystals we implement the computationally less expensive first method (out of the described three), switching off lattice relaxation at all stages and using the experimentally known positions of atoms in the crystal.

\begin{table}[]
\caption{Electronic (e) and ionic (i) contributions for the computed $\alpha_{zz}$ for all studied TMDs. $\omega_{0}$ corresponds to the optical frequency.}
\begin{tabular}{c|c|c|c}
      & $\alpha_{zz}^{e}(\omega > \omega_{0})\  [\mathrm{\AA}^3]$&  $\alpha_{zz}^{e}(\omega > \omega_{0})\  [\mathrm{\AA}^3]$ &  $\alpha_{zz}^{e+i}(\omega \rightarrow 0)\  [\mathrm{\AA}^3]$          \\\hline
MoS$_2$  &  44.40   &44.202 &   44.287                           \\
MoSe$_2$ &  52.295  &52.646 &   52.646                            \\
WS$_2$   &  43.947  &44.138 &   44.142                 \\
WSe$_2$  &  52.032  &52.507 &   52.507          \\
MoTe$_2$ &  67.897  &68.050 &   68.232           \\\hline
%hBN   &11.20        & 2.62      & 12.07        & 3.00 \\\hline
\end{tabular}
\label{tab:polarizability_elec_ionic}
\end{table}

\subsection{TMD bilayers}

In the analysis of TMD bilayers we take into account that those can be composed of qualitatively different stackings.
In one, known as 2H stacking (commonly synthesized in bulk crystals) the unit cells of consecutive layers are inverted as shown in Fig. \ref{Fig:Energy_versus_Displacement_field_sl_MoS2}(c).
In the other, which  corresponds to stacked consecutive layers in 3R-TMD polytypes, the consecutive layers have parallel orientation of the unit cells and alignment of chalcogen atom in one layer with a metal in the other as illustrated in Fig \ref{Fig:Energy_versus_Displacement_field_sl_MoS2}(c).
It has been recently shown\cite{Weston2020,Ferreira2021,Enaldiev_2021,Magorrian2021,Enaldiev2022,Li2017,Woods2021,Wallet_PRB_2021,Herrero_2021} that such bilayers exhibit interlayer charge transfer and a spontaneous out-of-plane ferroelectric polarization with the opposite orientation for XM and MX stacking configurations (see Fig \ref{Fig:Energy_versus_Displacement_field_sl_MoS2}(c)).
The results of the crystal energy computation, $U(D)$, for all of those configurations are shown in Fig. \ref{Fig:Energy_versus_Displacement_field_sl_MoS2}(d), using MoS$_2$ as typical example. 
We use such computed data to fit both spontaneous ferroelectirc polarization, $P$, and $\alpha_{zz}$ values using
a parabolic dependence which now incorporates a linear term $PD$ to account for the spontaneous interface electric dipole, 
\begin{equation}
U-U_0 = -\frac{PD}{\epsilon_0} - \frac{1}{2}\frac{\alpha_{zz}^{2L} D^2 }{ \epsilon_0 \mathcal{A}}.
\label{eq:total_energy_bilyer}
\end{equation}

\begin{table}[]
\caption{Values of $P$ and $\alpha_{zz}$  obtained by fitting DFT data to  Eq. (\ref{eq:total_energy}) and  (\ref{eq:total_energy_bilyer}) for monolayer and multilayer MoS$_2$ crystals respectively.
In the second column we indicate the interfaces that contribute to the ferroelectric prolarization (FP) of each crystal (see Fig. \ref{Fig:alpha_vs_Nlayers} and \ref{Fig:Energy_versus_Displacement_field_sl_MoS2}). 
In the fifth column we show the ratio between  $\alpha_{zz}$ and $\alpha_{zz}^{1L}$, where the latter corresponds to  $\alpha_{zz}$  of a monolayer.
}%, using a quadratic expression: $-bD-aD^2$.
%We note that $b=0$ for monolayers and for multi-layer crystals with inversion symmetry.
\label{tab_fittings}
\begin{tabular}{c|cccc}
\hline
\hline
 \multirow{2}{*}{MoS$_2$}  & FP interfaces  	   &  $P$   & $\alpha_{zz}$ & $\alpha_{zz} / \alpha_{zz}^{1L}$  \\ 
    		 &  &$ 10^{-4} e/\mathrm{\AA}   $ & $ \mathrm{A^3} $ &    \\
\hline
monolayer & -  & 0  &44.40  & 1.00 \\\hline
2H &-& 0      &88.81  & 2.00 \\ 
MX & $\boldsymbol{P}_{\mathrm{MX}}$ & -3.6         &88.91  & 2.00  \\ 
XM & $\boldsymbol{P}_{\mathrm{XM}}$   & 3.6        &88.89  & 2.00\\ \hline
2H-2H & - &0         &135.16 & 3.04 \\ 
XM-MX &  $\boldsymbol{P}_{\mathrm{XM}}+\boldsymbol{P}_{\mathrm{MX}}$  &0      &135.33 & 3.05 \\ 
XM-2H &  $\boldsymbol{P}_{\mathrm{XM}}$ &3.7      &135.24 & 3.05\\
XM-XM &  $2\boldsymbol{P}_{\mathrm{XM}}$ & 7.3  & 135.32 & 3.05 \\\hline
2H-2H-2H &   -  &0      &180.50 & 4.07 \\ 
XM-2H-MX &  $\boldsymbol{P}_{\mathrm{XM}}+\boldsymbol{P}_{\mathrm{MX}}$ &0    &180.56 & 4.07\\
2H-2H-XM & $\boldsymbol{P}_{\mathrm{XM}}$ &3.7  &180.57 & 4.07 \\ 
XM-2H-XM & $2\boldsymbol{P}_{\mathrm{XM}} $ &7.3   &180.67 & 4.07\\
\hline
\hline
\end{tabular}
\end{table}
The values that are extracted for the out-of-plane polarizability  of MoS$_2$ bilayers with all the described stackings coincide  (see Table \ref{tab_fittings}), with the DFT computation accuracy, and appear to be  approximately twice the value of monolayer polarizability.
This relation between the monolayer and bilayer polarizabilities is systematically reproduced for all other four studied TMDs.

\subsection{Trilayers and tetralayers of TMDs}

To test the scaleability of $\alpha_{zz}$ further we considered trilayers and tetralayers with various stacking interfaces (2H-2H, XM-MX, XM-2H, XM-XM, 2H-2H-2H, XM-2H-MX, 2H-2H-XM, XM-2H-XM).
The results of the DFT computed $U(D)$ depedence for each of those systems are displayed in  Fig. \ref{Fig:Energy_versus_Displacement_field_3l_4l_MoS2}(a) and (b) and analysed using Eq. (\ref{eq:total_energy_bilyer}).
This produces polarizabilty values which scale linearly with the number of layers for all of those configurations as listed in Table  \ref{tab_fittings}.
The values of $P$ obtained from the same data using Eq. (\ref{eq:total_energy_bilyer}) also correspond well to algebraic summation of independent contribution of consecutive interfaces which either compensate each other or double the resulting value (see Table  \ref{tab_fittings}), depending on the type of interfaces in the layer stacking illustrated in  Fig. \ref{Fig:Energy_versus_Displacement_field_3l_4l_MoS2}(c) and (d).
Overall the data for all five different TMDs are collected in Fig.\ref{Fig:alpha_vs_Nlayers}(a) where one can see that $\alpha_{zz}^{NL} = N \alpha_{zz}^{1L}$, where $N$ is the  number of layers.
%
%
%%%%%%%%%%%%%%%%%%%%%%%%%%%%%%%%%%%%%%%%%%%%%%
\begin{figure*}%[!h]
\includegraphics[width=1.0\linewidth ]{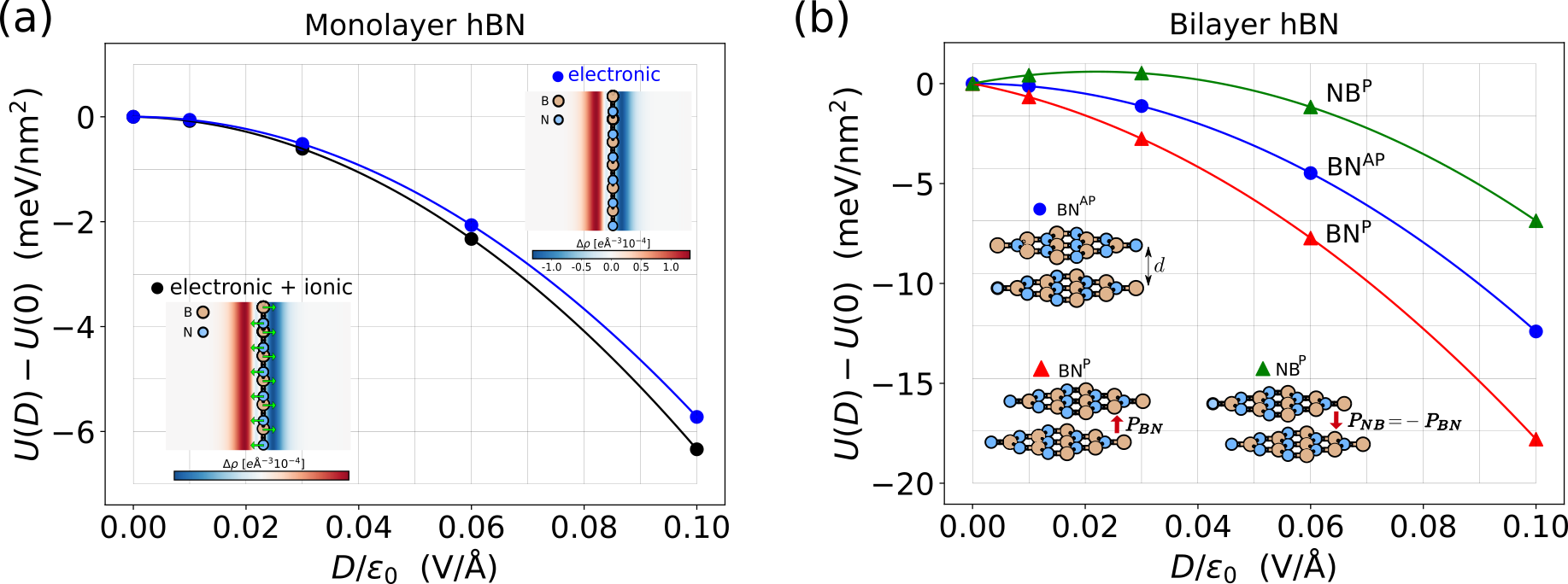}
\caption{ Total energy per unit cell area ($U$) of a monolayer of hBN (a) and bilayer of hBN with BN$^\mathrm{AP}$, BN$^\mathrm{P}$ and NB$^\mathrm{P}$-stacking configurations (b) as a function of  an external out-of-plane displacement field ($D$).   $U(0)$ is the total energy per unit cell area for a $D=0$. 
Symbols represent DFT data, whereas solid lines represent the fittings to DFT data, which were computed by using the expression in Eq. (\ref{eq:total_energy}). 
%A large out-of-plane periodicity ($L_{z} = 80\  \mathrm{\AA}$) between the crystals was used in our calculations (see Methods). 
The out-of-plane averaged charge density difference $\Delta \rho$ for a monolayer is shown as an inset in (a), where blue and red regions represent the accumulation and depletion of electrons, respectively.
The different stacking configurations used for our bilayers of hBN are shown in (b), where $d$ is the interlayer distance and we show the directions of the ferroelectric polarization $\boldsymbol{P}_{\mathrm{NB}}$ and $\boldsymbol{P}_{\mathrm{BN}}$ due to interlayer charge transfer. The parameters obtained from our fittings are gathered in Table \ref{tab_fittings_hBN}.
}
\label{Fig:Energy_versus_Displacement_field_sl_hBN}  
\end{figure*}
%%%%%%%%%%%%%%%

\subsection{Monolayer and multi-layer hBN crystals}

In this subsection we repeat the analysis of  $\alpha_{zz}$ for hexagonal boron  nitride crystals, this includes the study of lattice relaxation and ionic contribution towards polarizability and additivity of ferroelectric polarization due to the interlayer charge transfer in inversion-asymmetric interfaces.
We carried out DFT calculations using the plane-wave code implemented in Quantum Espresso package \cite{QE1,QE2}. 
A plane wave cut-off of 90 Ry was used for all calculations with hBN films, where the integration over the Brillouin zone was performed using the scheme proposed by Monkhorst-Pack \cite{kgrid} with a grid of $9 \times 9\times 1$.
For hBN calculations, we used norm conserving pseudopotentials and an exchange correlation functional that is approximated by using the PBE method\cite{PBE},  with the inclusion of a van-der-Waals functional described by Tkatchenko–Scheffler vdW-TS\cite{Tkatchenko2009}. 
The convergence threshold for self-consistency was  set to $10^{-9}$ Ry. 
A $z$-dependent sawtooth potential was used to induce an out-of-plane displacement field where an out-of-plane Coulomb truncation \cite{CoulombCut} was considered.
Relaxation calculations for the atomic coordinates were done using the BFGS quasi-newton algorithm, with a threshold convergence for the total force acting on the atoms less than $10^{-5}$ Ry/Bohr

In contrast to TMDs the out-of-plane relaxation of ions (for a lattice constant fixed to the experimentally known value) gives two clearly distinguishable $U(D)$ dependence which result in the values of $\alpha_{zz}^{e}$
 and $\alpha_{zz}^{e+i}$ listed in Table \ref{tab_fittings_hBN}.
An example of calculations for three stacking configurations of hBN bilayers (with parallel and anti-parallel unit cell orientations), performed with lattice relaxation implemented in the code, are displayed in Fig.  \ref{Fig:Energy_versus_Displacement_field_sl_hBN}(b).
 The difference between the $U(D)$ dependence results from the ferroelectric polarization of bilayers with parallel unit cells orientations\cite{Li2017,Woods2021,Walet2021_2,Herrero_2021}.
 The values of $\alpha_{zz}^{e+i}$ for these bilayers as well for trilayers and tetralayers listed in Table \ref{tab_fittings_hBN} demonstrate perfect scaleability with the number of layers, $\alpha_{zz}^{NL} = N\alpha_{zz}^{1L}$, for both electronic and combined electronic plus ionic responses. Moreover, ferroelectric polarization exhibits additivity of contribution of individual interfaces. 
\begin{table*}
\caption{Values of $P$ and $\alpha_{zz}$  obtained by fitting DFT data to  Eq. (\ref{eq:total_energy})  for monolayer (1L) and multi-layer (2L, 3L, 4L) hBN  crystals respectively.
In the second column we indicate the interfaces that contribute to the ferroelectric prolarization (FP) of each crystal (see Fig. \ref{Fig:Energy_versus_Displacement_field_sl_hBN}). 
In the sixth and ninth column we show the ratio between  $\alpha_{zz}$ and $\alpha_{zz}^{1L}$ with and without ionic contribution.
}
\label{tab_fittings_hBN}
 \begin{tabular}{cc|c|ccc|ccc}
\hline
& & & \multicolumn{3}{c}{Electronic (e)} & \multicolumn{3}{c}{Electronic (e) + ionic (i)} \\
\hline
& \multirow{2}{*}{hBN}  & FP interfaces  	   &  $P$   & $\alpha_{zz}^e$ & $\alpha_{zz}^{e} / \alpha_{zz}^{1L (e)}$ &  $P$   & $\alpha_{zz}^{e+i}$ & $\alpha_{zz}^{e+i} / \alpha_{zz}^{1L (e+i)}$  \\ 
    		 & &  &$ 10^{-4} e/\mathrm{\AA}   $ & $ \mathrm{A^3} $ &   &  $ 10^{-4} e/\mathrm{\AA}   $ & $ \mathrm{A^3} $ &     \\
\hline
1L & & -  & 0  &11.20  & 1.00 &  0 & 12.07 & 1.00 \\\hline
\multirow{3}{*}{2L} & BN$^{\mathrm{AP}}$ &-& 0      &22.56  & 2.01 & 0  & 24.23  & 2.01 \\ 
& BN$^{\mathrm{P}}$ & $\boldsymbol{P}_{BN}$ & 5.8        &22.46  & 2.01 & 5.5  & 24.09  & 2.00  \\ 
& NB$^{\mathrm{P}}$ &  $\boldsymbol{P}_{NB}$  & -5.7        &22.47  & 2.01 & -5.4  & 24.09  & 2.00\\ \hline
\multirow{3}{*}{3L} & BN$^{\mathrm{AP}}$-BN$^{\mathrm{AP}}$& - &0         &33.88 & 3.02 &0         &36.32 & 3.01 \\ 
& BN$^{\mathrm{P}}$-NB$^{\mathrm{P}}$  &  $\boldsymbol{P}_{\mathrm{BN}}+\boldsymbol{P}_{\mathrm{NB}}$  &0      &33.67 & 3.01 &0      &36.03 & 2.99\\
& BN$^{\mathrm{P}}$-BN$^{\mathrm{P}}$  &  $2\boldsymbol{P}_{\mathrm{BN}}$  &12.0      &33.69 & 3.01 &11.7      &36.19 & 3.00\\ \hline
\multirow{3}{*}{4L} & BN$^{\mathrm{AP}}$-BN$^{\mathrm{AP}}$-BN$^{\mathrm{AP}}$ &   -  &0      & 45.17 & 4.03 &0     &48.73 & 4.04\\ 
& BN$^{\mathrm{P}}$-NB$^{\mathrm{P}}$-BN$^{\mathrm{P}}$ & $2\boldsymbol{P}_{\mathrm{BN}}+\boldsymbol{P}_{\mathrm{NB}}$ &5.9   &44.88 & 4.01 &6.3   &49.05 & 4.06\\
& BN$^{\mathrm{P}}$-BN$^{\mathrm{P}}$-BN$^{\mathrm{P}}$ & $3\boldsymbol{P}_{\mathrm{BN}}$ &17.5   &44.86 & 4.01 &17.9   &48.48 & 4.02\\
\hline
\hline
\end{tabular}
\end{table*}
\begin{table*}
	\caption{Dielectric constant $\epsilon_{zz}$ of bulk TMDs studied in this work and hBN, monolayer band gap $E_g^{1L}$,  monolayer out-of-plane polarizabilites  $\alpha_{zz}^{1L}$.  For the latter we show the value calculated here,  theoretical values from other works that carried out DFPT (density functional perturbation theory) calculations, other theoretical methods different from DFPT and measured values obtained experimentally.  \label{tab_polarizability}}
	\begin{tabular}{l|c|c|c|c||c|c}
		\hline
		\hline
&   This work & DFPT & Other methods & Measured & $E_g^{1L}$ [eV] &  $\alpha_{zz}$ [$\mathrm{\AA}^3$]  \\\hline
$\epsilon^{\mathrm{MoS}_2}$   &   6.10      & 7.6\cite{Liang2014}, 6.87\cite{Pike2018}, 6.9\cite{Laturia2020} & 3.92\cite{Ghosh2013}, 6.3\cite{Koo2017}, 10.08\cite{Farkous2019} &  4.9\cite{Neville1976}, 3.7\cite{Lu2014}\footnote{Layer with a thickness of 30 nm.}, 6.2\cite{Wieting1971}  & 1.71  & 44.47   \\
$\epsilon^{\mathrm{WS}_2}$    &  5.84       &   6.87\cite{Liang2014}, 6.4\cite{Laturia2020}  & 5.14\cite{Ghosh2013}, 6.0\cite{Koo2017}, 6.39\cite{Farkous2019} & & 1.67  & 43.95 \\
$\epsilon^{\mathrm{MoSe}_2}$  &  7.34       & 8.5\cite{Laturia2020} & 6.07\cite{Ghosh2013},7.8\cite{Koo2017} & & 1.41  & 52.29\\
$\epsilon^{\mathrm{WSe}_2}$   &  7.20       & 7.8\cite{Laturia2020} & 5.16\cite{Ghosh2013},7.7\cite{Koo2017} & & 1.33  & 52.03  \\
$\epsilon^{\mathrm{MoTe}_2}$  &  10.69      & 16.10\cite{Ghosh2013} & 10.4\cite{Laturia2020} & & 1.02  & 67.91 \\
$\epsilon^{\mathrm{hBN}}_{(\omega > \omega_{0})}$     &  2.64       &   3.38 \cite{Ohba2001}, 3.57\cite{Cai2007}, 5.09\cite{Wang2016}, 3.76\cite{Laturia2020} & 2.7\cite{Koo2017} &  5.09 \cite{Geick1966},  3.5-3.8\cite{Ahmed2018}, 3 to 5\cite{Kim2015}) &  4.62  & 11.24 \\
$\epsilon^{\mathrm{hBN}}_{(\omega \rightarrow 0)}$          &  3.00       &   2.84 \cite{Ohba2001}, 2.95\cite{Cai2007}, 3.61\cite{Wang2016}, 3.03\cite{Laturia2020} &  &  4.10 \cite{Geick1966}& 4.62  & 12.07\\

		\hline
		\hline
		%\multicolumn{9}{l}{$^*)$ For $\alpha^{\rm 3R}_{zz}$, $\chi$ and $\epsilon_{zz}$ we show the values obtained with}
	\end{tabular}
	\label{Tab:table_dielectric}
\end{table*}

\section{Discussion}
In order to recalculate the computed $\alpha_{zz}$ into dielectric constant of a medium composed of many layers in a TMD or hBN crystal, we use the following expression
\begin{equation}
\epsilon_{zz} = \left(1-\frac{\alpha_{zz}^{1L}}{ \mathcal{A} d} \right)^{-1}.
\label{eq:dielectric_zz}
\end{equation}
which has been successfully implemented before in the analyses of layered materials\cite{Tian2020,Slizovskiy2021,Enaldiev2022}.
In Table \ref{Tab:table_dielectric}  we gather the resulting values of $\epsilon_{zz}$ (obtained from the corresponding computed values of  $\alpha_{zz}^{1L}$) for all the materials studied here). 
As we found in section II that for TMDs the ionic contribution is negligibly small, the results of $\epsilon_{zz}$  are expected to be the same for both zero and high frequencies.
For hBN the low and high frequency dielectric constants are distinguishable due to a substantial contribution of ions towards static polarizability evident in Table \ref{Tab:table_dielectric}.
Nevertheless the overall dielectric constant of hBN is smaller than of TMDs due to a weaker electronic polarizabilty which we attribute to a much larger band gap of this material. %
\section{Conclusions}

The analysis presented here demonstrates linear scaling of polarizability of multilayer TMDs and hBN with the number of layers of these van der Waals materials which suggests that layers respond to the out-of-plane perturbation independently of each other. 
This enabled us to quantify the dielectric constant of these materials determined by such polarizabilty of the layers with the values of the computed polarizabilities that we compare with the results of previous computations and some experimentally available data in Table \ref{Tab:table_dielectric}.
\section*{Acknowledgments}
This work was supported by EC-FET European Graphene Flagship Core3 Project, EC-FET Quantum Flagship Project 2D-SIPC, EPSRC grants EP/S030719/1 and EP/V007033/1, and the Lloyd Register Foundation Nanotechnology Grant.
Computational resources were provided by the Computational Shared Facility of the University of Manchester and the ARCHER2 UK National Supercomputing Service (https://www.archer2.ac.uk) through EPSRC Access to HPC project e672.

\newpage

\bibliography{sample}

\end{document}